# Low-spatial coherence electrically-pumped semiconductor laser for speckle-free full-field imaging

B. Redding[1], A. Cerjan[1], X. Huang[2], M. L. Lee[2], A. D. Stone[1], M. A. Choma[1,3,4,5], and H. Cao[1*]

[1]*Department of Applied Physics, Yale University, New Haven CT 06520*
[2]*Department of Electrical Engineering, Yale University, New Haven CT 06520*
[3]*Department of Diagnostic Radiology, Yale School of Medicine, New Haven CT 06520*
[4]*Department of Biomedical Engineering, Yale University, New Haven CT 06520*
[5]*Department of Pediatrics, Yale School of Medicine, New Haven CT 06520*
*\*hui.cao@yale.edu*

**Abstract**

The spatial coherence of laser sources has limited their application to parallel imaging and projection due to coherent artifacts, such as speckle. In contrast, traditional incoherent light sources, such as thermal sources or light emitting diodes (LEDs), provide relatively low power per independent spatial mode. Here, we present a chip-scale, electrically-pumped semiconductor laser based on a novel design, demonstrating high power per mode with much lower spatial coherence than conventional laser sources. The laser resonator was fabricated with a chaotic, D-shaped cavity optimized to achieve highly multimode lasing. Lasing occurs simultaneously and independently in ~1000 modes, and hence the total emission exhibits very low spatial coherence. Speckle-free full-field imaging is demonstrated using the chaotic cavity laser as the illumination source. The power per mode of the sample illumination is several orders of magnitude higher than that of a LED or thermal light source. Such a compact, low-cost source, which combines the low spatial coherence of a LED with the high spectral radiance of a laser, could enable a wide range of high-speed, full-field imaging and projection applications.

**Introduction**

Spatial coherence is a defining characteristic of laser emission. High spatial coherence allows focusing electromagnetic energy to a small spot or collimation of an optical beam over a long distance. However, spatial coherence can also introduce coherent artifacts such as speckle, since any uncontrolled scattering in the imaging system can cause multi-path interference. These artifacts have limited the use of lasers in full-field imaging applications ranging from traditional wide-field microscopes to laser projectors and photolithography systems. Instead, traditional low spatial coherence sources such as thermal light sources and LEDs are still used for illumination in most full-field imaging applications, despite having lower power per mode, poor collection efficiency, and less spectral control than lasers. These limitations are particularly pronounced in applications requiring high-speed imaging, or imaging in absorbing or scattering media, prompting the use of raster-scanning based laser imaging systems. For parallel imaging and projection applications, the ideal illumination source would combine the high power per mode of a laser with the low spatial coherence of an LED. The common approach to achieve this combination is by reducing the effective spatial coherence of a traditional laser using, e.g. a spinning diffuser[1], a colloidal solution [2], or a MEMs mirror[3]. However, these techniques require averaging over many speckle patterns in time, mitigating the advantage of using bright sources such as lasers or superluminescent diodes for high-speed imaging applications.

An alternative and more effective approach is to design a laser that generates light in an intermediate number of spatial modes, producing sufficiently low spatial coherence to suppress speckle, while maintaining higher power per mode than an LED or a thermal light source. To this end, we recently demonstrated that a random laser[4,5] and a degenerate laser[6] can be engineered to efficiently control the spatial coherence of emission. However, the required optical pumping, together with the large size and cost of these systems, may limit their use outside of a research setting. Similar functionality has also been realized by combining many independent lasers to synthesize a low-spatial coherence laser source. For example, a recently developed vertical cavity surface emitting laser (VCSEL) array combines ~1000 mutually incoherent lasers on a single chip[7]. However, this approach requires a relatively large chip since the lasers must be sufficiently separated to remain uncoupled. In addition, the individual VCSELs typically lase in a single longitudinal mode due to the short cavity length, limiting the ability to generate broadband emission as required for imaging modalities such as optical coherence tomography (OCT) and laser ranging.

In this work, we demonstrate a chip-scale electrically-pumped semiconductor laser that produces intense emission with low spatial coherence. Specifically, we designed a chaotic microcavity to support highly multimode lasing, and experimentally realized lasing in ~1000 mutually incoherent modes in a single cavity. The lasing emission was used as an illumination source for speckle-free full-field imaging. The chaotic cavity laser was compared to a standard Fabry-Perot (FP) cavity laser which produced emission with high spatial coherence. The chaotic microcavity laser was fabricated by photolithography and wet etching, and hence was relatively simple and compatible with mass production. The lasing performance was robust against cavity shape deformation and boundary roughness. Such a compact, low-cost laser with low spatial coherence and high radiance could enable a wide range of high-speed, full-field imaging and projection applications.

**Results**

**Chaotic cavity design.** To achieve multimode lasing, the cavity must support a large number of resonances with similar quality ($Q$) factor so that their lasing thresholds are comparable. Moreover, the competition for gain should be minimized so that the first few lasing modes do not deplete the gain and prevent other modes from lasing. A promising approach to satisfying these requirements was to employ a two-dimensional "chaotic cavity" design. The term refers to planar dielectric cavities in which the ray dynamics are chaotic over all or much of the phase space[8,9]. Such cavities do not lead to chaotic laser *dynamics* (as has been realized in other types of laser systems[10]). Instead the presence of ray chaos leads to many pseudo-random spatial modes that, on average, fill the entire cavity uniformly (with independent Gaussian fluctuations on the wavelength scale). In such a cavity there will be no subset of strongly-preferred high Q modes, and a large enough cavity of this type will support many modes with similar thresholds.

In this work, we used a fully chaotic cavity consisting a disk of radius, $R$, with a section removed along a chord with length parameterized by $r_0$ as shown in Fig. 1(a), referred to as a "D-shaped cavity". This D-shaped geometry is known to support chaotic ray dynamics[11,12]; and has been employed previously to improve the pumping efficiency of fiber amplifiers[13]. Compared to an on-chip random laser that consists of numerous subwavelength scatterers (e.g., air holes)[14], the chaotic microcavity laser is not only simpler and less expensive to fabricate, but also has lower loss from out-of-plane scattering and non-radiative recombination of carriers at the semiconductor/air interface.

By introducing the flat cut, the high-$Q$ whispering gallery modes (WGM) of a circular disk are eliminated. If one neglects outcoupling loss and treats the system as an ideal "billiard", then, for $0 < r_0 < R$, it has been shown[8,9] that the ray motion is fully chaotic, meaning that there exist no stable periodic ray orbits. Quantizing the wave equation on such a domain will lead to many chaotic modes, spread over the full cavity, with similar $Q$ values, and hence similar lasing threshold. Although the system is fully chaotic for all values of $r_0$, as the value of $r_0$ approaches zero or $R$, the cavity approaches the non-chaotic limit (of a full or semi-circle) and exhibits weaker instability in the motion leading to wave solutions which are much less random and uniform in space. Thus, one would expect the optimal value for generating states with similar $Q$ values is $r_0 = R/2$.

However narrowing the distribution of lasing thresholds by designing a cavity to have similar $Q$ values is only one aspect of maximizing the number of lasing modes. One must also reduce the mode competition which prevents modes with somewhat lower $Q$ values from turning on once the first few modes are lasing. One cannot evaluate this effect using only $Q$-values and a linear analysis; it is determined by non-linear cross-gain saturation and spatial hole burning in the active cavity. Since chaotic modes overlap in space it is not immediately clear that mode competition is minimized by choosing the geometry with the most chaotic wave solutions.

To address this question theoretically we use Steady-state *Ab initio* Laser Theory (SALT), a relatively new approach, which treats the cavity geometry and modal interactions exactly, assuming that a stable multimode steady-state exists[15–17]. Specifically, we investigate the effect of mode competition on the lasing thresholds as a function of cavity shape, and compare the results to the expected thresholds in the absence of mode competition [Fig. 1 (e)]. First, we investigate the effect of the $Q$-value distribution alone, without including mode competition. The dashed line connecting the data points in Fig. 1(e) are the *non-interacting pump* thresholds for the first ten lasing modes calculated for a circular microdisk and three D-shaped cavities with varying values of $r_0$. For each cavity shape, the pump was normalized to be unity at the first threshold. As expected, the lowest (normalized) thresholds for each mode (n=2,3…10) occurs for $r_0 = R/2$; this reflects the narrow distribution of $Q$-values in the most chaotic cavity shape. In contrast, e.g. for the circular cavity, there are two high-$Q$ whispering gallery modes which turn on at nearly the same pump value and then the next modes within the gain curve have much lower $Q$ and require much higher relative pumps to reach threshold.

Next, we present the full non-linear calculation at a pump value corresponding to five modes lasing, shown by the data points joined by the solid lines. This demonstrates the additional effect of mode competition. Once the laser is significantly above threshold, so that the effects of gain saturation are strong, all shapes have their corresponding thresholds pushed to higher pump values, reducing the number of modes lasing at a given pump. The difference between the dashed and solid lines represents the effect of mode competition. The simulations clearly show that the most chaotic cavity, with $r_0 = R/2$, has the weakest mode competition and the largest number of modes lasing for the same normalized pump. Figs 1(a-c) illustrates the qualitative reason for this behavior. For the less chaotic shapes ($r_0$=0.3, 0.7), the lasing modes are still somewhat localized in space, apparently near some unstable periodic orbits closer to the boundary. The spatial localization of these modes makes their mode competition stronger, clamps the gain, and makes it difficult for many modes to lase simultaneously. The extreme case, of course, is the circular cavity (not shown), for which the high-$Q$ WGMs are all highly-peaked close to the cavity boundary, and the fourth lasing modes turns on at 9.5 times the 1st threshold pump, whereas the D-shaped cavity laser with $r_0 = R/2$ has a fourth mode threshold only 2% higher than the first threshold. Note that the cavities we simulated are much smaller than those we fabricated, due to numerical constraints on two-dimensional calculations of this type. In the larger cavity sizes used in our experiments many more modes are expected to lase simultaneously for the same pump and shape. Nevertheless, the simulations were a valuable guide for determining the optimal cavity design, as confirmed by the experiments.

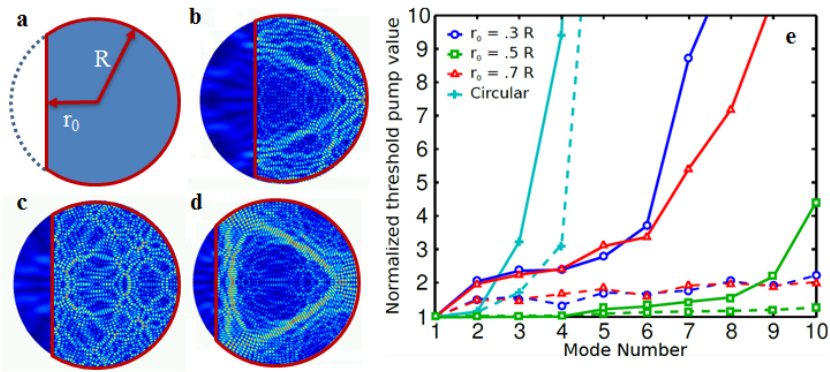

**Figure 1 | Design and simulation of D-cavity lasers to optimize multi-mode lasing.** (a) A schematic of the D-shaped cavity that consists of a circle with a flat edge at a distance $r_0$

from the center. (b-d) Numerically calculated (false color) amplitude of the electric field distribution for the highest $Q$ mode in cavities of radius $R = 5$ μm and $r_0 = 0.3R$, $0.5R$, and $0.7R$, respectively. The modes have transverse-magnetic (TM) polarization with the electric field perpendicular to the cavity plane. The mode in the cavity of $r_0 = 0.5R$ displays the most uniform spatial intensity distribution. (e) Calculated pump thresholds of the first ten lasing modes in four cavities of $r_0 = R$ (circle), $0.7R$, $0.5R$, and $0.3R$, normalized to the threshold of the first lasing mode. Dashed lines: non-interacting thresholds reflecting the $Q$-value distribution of the passive cavity. The $r_0 = 0.5R$ cavity has the narrowest distribution of $Q$-values, which is favorable for multimode lasing at relatively low pump values. Solid lines: interacting thresholds, calculated from SALT up to the fifth lasing threshold, and extrapolated using the SPA-SALT approximation [17] up to the tenth lasing threshold. Mode competition causes the actual thresholds to increase compared to the non-interacting estimates, but has the weakest effect for the $r_0 = 0.5R$ cavity shape (due to reduced localization of the lasing modes as discussed in the text). Hence the $r_0 = 0.5R$ cavity is optimal for maximizing the number of lasing modes.

**Experimental realization of the D-cavity laser.** We then fabricated the D-shaped cavity with the optimized geometry of $r_0 = 0.5R$. To realize lasing with electrically pumping, we used a commercial laser diode wafer consisting of a P-I-N junction with gain provided by a GaAs quantum well. D-shaped cavities with radius $R$ ranging from 100 μm to 500 μm were fabricated using standard photolithography and wet etching techniques and characterized using the experimental setup shown schematically in Fig. 2(a). The optical image of a D cavity of $R = 500$ μm is shown in Fig. 2(b).

At low pump current, the electroluminescence spectrum of the quantum well was broad and smooth [red line in Fig. 2(c,d)]. As the pump current increased, the emission spectrum narrowed due to amplification of spontaneous emission (ASE). When the current exceeded a threshold value, discrete narrow peaks appeared in the emission spectra of smaller D cavities (e.g. $R = 100$ μm or 250 μm), as seen in Fig. 2(c). The full-width-at-half-maximum (FWHM) of these peaks is less than 0.1 nm. The intensity of these peaks grew rapidly with pump current. These results indicate the onset of lasing action. The number of lasing peaks increased with the cavity size as expected, and merged to a continuous band in the cavities of $R = 500$ μm [blue line in Fig. 2(d)]. The spectrally integrated emission intensity grew superlinearly with the current [blue circles in Fig. 2(e)], while the emission bandwidth was reduced [red squares in Fig. 2(e)]. These data provide further evidence for lasing. Note that the L-I curve did not exhibit an abrupt transition, because the measured emission was from a large number of modes with slightly different lasing thresholds. We also tested lasing in a 100 μm wide and 2 mm long FP cavity. The FP laser displayed a sharp threshold in the L-I curve and a narrow emission spectrum with FWHM ~ 0.3 nm above threshold.

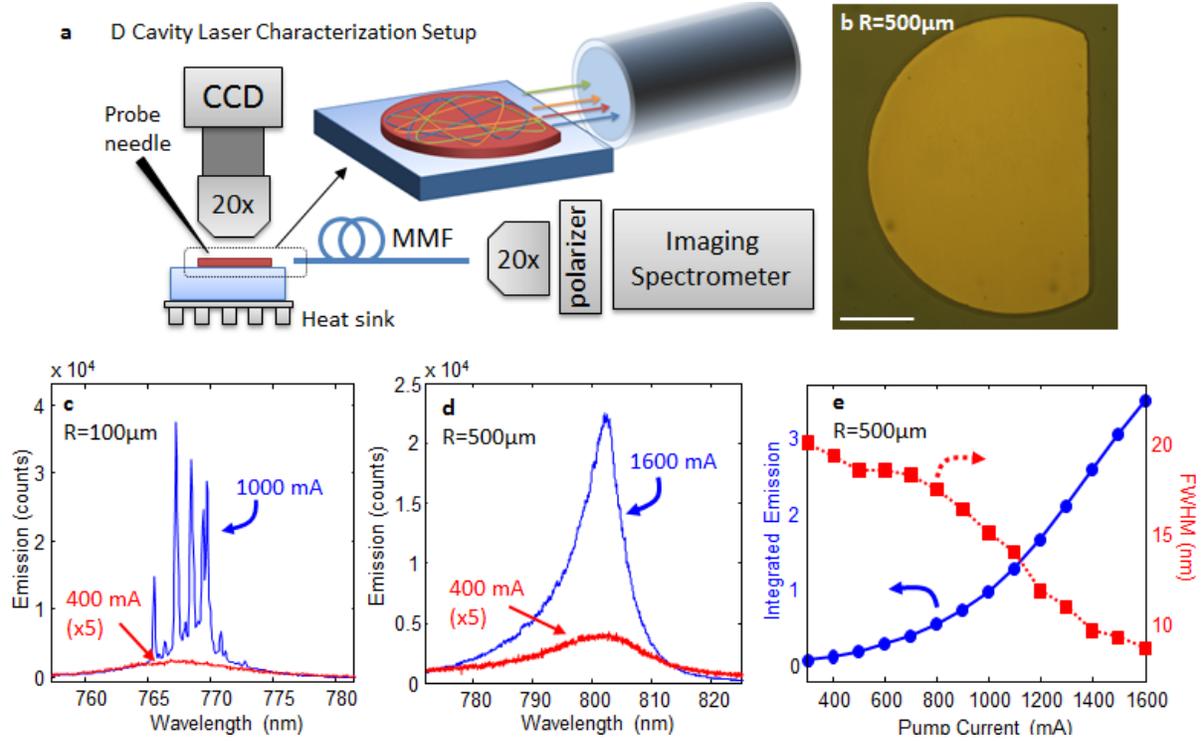

**Figure 2 | Experimental realization of the D-cavity laser.** (a) Schematic of the experimental setup to characterize the D cavity laser. A microscope is used to image the chip from above to align the probe needle to a single cavity. Electrical current is injected through the needle to the cavity, and the emission is collected from the edge of the cavity with a multimode fiber as shown in the inset. The end facet of the multimode fiber is then imaged by a 20x objective onto the entrance slit of an imaging spectrometer through a linear polarizer. (b) Optical microscope image of the $R = 500$ μm D cavity laser taken from above. The scale bar is 250 μm. (c) The emission spectra collected from a D cavity laser with $R=100$ μm and $r_0 = 0.5R$ at a pump current of 400 mA (red line) and 1000 mA (blue line). The emission spectrum collected at 400 mA was magnified 5× for clarity. At 1000 mA, discrete narrow peaks appear in the emission spectrum. (d) The emission spectra collected from a D cavity laser with $R = 500$ μm and $r_0 = 0.5R$ at a pump current of 400 mA (red line) and 1600 mA (blue line). A dramatic narrowing of the emission spectrum was observed at high pump current. (e) Spectrally integrated emission intensity and emission bandwidth as a function of the pump current for the D cavity with $R = 500$ μm and $r_0 = 0.5R$. As the emission intensity grows non-linearly with the pump current, the emission bandwidth drops quickly.

**Spatial coherence characterization.** Next we characterized the spatial coherence of the D cavity laser and the FP laser by collecting emission with a 1-meter long, step-index multimode fiber (core diameter = 105 μm, numerical aperture = 0.22) and measuring the speckle contrast at the end of the fiber. Speckle is formed in a multimode fiber by interference of the guided modes and the speckle contrast is defined as $C=\sigma_I/<I>$, where $\sigma_I$ is the standard deviation of the intensity and $<I>$ is the average intensity. Since the chaotic lasing modes in the D cavities have different emission patterns, they will excite the fiber modes with different relative phase and amplitude, thus generating distinct speckle patterns. If these modes lase independently then their emission is mutually incoherent and they do not interfere, causing their speckle patterns to add in intensity. Thus the speckle contrast, $C$, decreases as $C=M^{-1/2}$, where $M$ is the number of uncorrelated speckle patterns[18]. Hence, from the speckle contrast at the end of the multimode fiber, we can estimate the number of independent lasing modes in the cavity. Since polarization mixing in the fiber could also reduce the speckle contrast, a linear polarizer was placed in front of the camera to eliminate this effect.

An optical image of the speckle pattern at the output facet of the multimode fiber by emission from the FP laser is shown in Fig. 3(a). This speckle pattern was recorded during a 1 μs interval of current injection; however the speckle pattern remained unchanged over multiple intervals, indicating that the same lasing modes were excited each time. The FP laser emission produced high contrast speckle, revealing high spatial coherence. The measured speckle contrast was $C = 0.58$, implying that the FP laser supported only ~3 independent transverse modes. Although the FP cavity was made of a 100 μm wide ridge waveguide that supports ~450 transverse modes, strong mode competition in this cavity limited lasing to just a few transverse modes. In contrast, the D cavity laser emission produced speckle with very low contrast, as seen in Fig. 3(b). The measured contrast of $C = 0.03$ suggested that ~1000 modes lased simultaneously and independently in the D cavity, as a result of suppression of mode competition.

The low speckle contrast in Fig. 3(b) confirms that the D cavity laser produced emission with low spatial coherence suitable for full-field imaging applications. However, a multimode fiber may reduce the effective spatial coherence of broadband emission[19], if the linewidth of individual lasing modes exceeds the spectral correlation width of the fiber[20]. Our estimation of the number of independent lasing modes assumed that the spectral width of each lasing mode was smaller than the spectral correlation width of the fiber (~0.1 nm), consistent with the observation of narrow lasing peaks of width < 0.1 nm from smaller D cavities. In the FP laser, there could be numerous longitudinal modes that correspond to a single transverse mode, but all these longitudinal modes would have the same emission pattern and create identical speckle in the fiber. Hence, only different transverse modes generated distinct speckle patterns and reduced the speckle contrast.

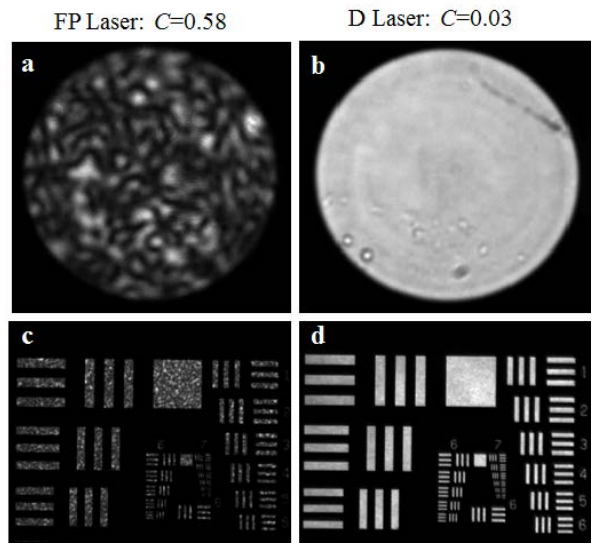

**Figure 3 | Spatial coherence of the D cavity laser and application to full-field imaging.** (a,b) Optical images of the speckle pattern at the end of a 1 meter long multimode fiber by emission from the FP laser (a) and D cavity laser (b). The FP laser emission produced high contrast speckle ($C = 0.58$), indicating that lasing was limited to ~3 transverse modes, whereas the D cavity laser produced low contrast speckle ($C = 0.03$), implying independent lasing in ~1000 modes. (c,d) The emission from the end of the multimode fiber was used to illuminate an Air Force resolution test chart in transmission mode through an immobile ground glass diffuser. The high spatial coherence of the FP laser produced a speckled image

(c) whereas the low spatial coherence of the D laser emission gave a high quality image with low speckle contrast (d).

**Speckle-free imaging.** Finally, we demonstrated that the D cavity lasers could be used as an illumination source for full-field speckle-free imaging. The laser output from the sidewall of the D-cavity provides a line emission that is coherent in the direction perpendicular to the cavity plane and incoherent in the direction parallel to the cavity plane. Such a line source may be directly applied to line scanning in parallel OCT or confocal imaging. Alternatively, we can make a 2D incoherent light source by coupling the laser emission to a multimode fiber. The fiber mixes the degrees of spatial coherence in different directions, making the degree of spatial coherence (or incoherence) isotropic. We used the output from the multimode fiber to illuminate an Air Force resolution test chart in transmission mode through an immobile ground glass diffuser (Thorlabs DG10-600-MD). Figure 3(c,d) are the images of the Air Force test chart illuminated by the FP laser and the D cavity laser. The speckle, clearly seen with the FP laser illumination, became invisible with the D cavity laser illumination.

In addition to providing speckle-free illumination, the D cavity laser produces much higher power per mode than traditional low spatial coherence sources such as thermal light sources and LEDs. This is because the D cavity laser concentrates emission in ~1000 spatial modes, whereas thermal sources and LEDs, with the spatial coherence length on the order of a wavelength, distribute light in many more spatial modes. From an imaging perspective, thermal light sources and LEDs are very inefficient, because only ~1000 modes are required to suppress speckle below the level observable by humans [21]. As a quantitative comparison, we estimated the photon degeneracy of the D laser, a measure of the number of photons per coherence volume. The photon degeneracy parameter $\delta = (P\ \delta z)/(h\nu\ c\ M)$, where $P$ is the emission power, $\delta z$ is the temporal coherence length, $h\nu$ is the photon energy, $c$ is the speed of light, and $M$ is the number of spatial modes[22]. The emission bandwidth of the $R = 500$ μm D cavity laser is 8 nm, giving $\delta z = 35$ μm. The output from the D cavity laser is nearly isotropic in the plane of the disk, and the multimode fiber collected approximately 15% of the total emission, which was ~4 mW during the 1 μs pump pulse. The estimated value of $\delta$ is ~ $10^2$, which is 4-5 orders of magnitude higher than that of a thermal source (at 4000 K $\delta$ ~$10^{-3}$)[22] and of a high efficiency LED ($\delta$ ~$10^{-2}$)[23]. The greatly improved photon degeneracy could be used to achieve much higher imaging speeds in light-limited full-field imaging applications. Moreover, we expect that with proper packaging and thermal management, the D cavity laser will operate in continuous wave (CW) at even higher power. Also, it may be possible to improve the emission collection efficiency by optimizing the chaotic cavity shape for directional emission or adding a reflecting boundary along the circular edge of the D cavity. In addition, the pulsed operation of D cavity lasers may be used for speckle-free stroboscopic imaging [24].

**Discussion**

Finally, we compare the D cavity laser to superluminescent diodes (SLD) based on amplified spontaneous emission (ASE), which share some similar characteristics. For example, both the D cavity laser and SLDs exhibit a superlinear increase of emission intensity with pump and produce relatively broadband emission. However, SLDs maintain relatively high spatial coherence, and cannot be used in full-field imaging applications without producing speckle. This is because the SLDs are typically realized in a ridge waveguide geometry where gain competition suppresses the amplification of high-order transverse spatial modes. The key distinction between a SLD and a traditional laser is that SLDs exhibit low temporal coherence (and hence broadband emission). This has made SLDs a popular light source for applications such as OCT which require broadband emission with high power per mode. By contrast, the D cavity laser achieves both low temporal coherence and low spatial coherence. Moreover, from a full-field imaging perspective, the distinction between ASE and laser emission is less important than achieving low spatial coherence while maintaining high photon degeneracy. The D cavity laser matches these requirements and illustrates the potential for tailoring the spatial and temporal coherence of lasers for target application. In the future, we will explore multiple quantum wells with varying thickness and/or composition to obtain broadband gain[25] which could enable a D cavity laser with even lower temporal coherence and a broader emission spectrum.

In summary, we have demonstrated a chip-scale, electrically-pumped semiconductor laser that combines low spatial coherence and high power per mode. The chaotic cavity shape was selected and optimized to enable highly multimode lasing. We fabricated the D cavity laser with standard photolithography and wet chemical etching. It supported ~1000 independent lasing modes, combining to produce low spatial coherence emission. We confirmed that the laser emission could be used as an illumination source for speckle-free full-field imaging. By designing the laser to support enough spatial modes to suppress speckle, without the over-abundance of modes present in traditional low spatial coherence sources, much higher power per mode was reached. Such a compact, low-cost and bright source with low spatial coherence could enable a wide range of high-speed, full-field imaging and ranging applications.

## Methods

### Wafer Structure

The D cavity laser was fabricated using a commercial laser diode wafer consisting of a P-I-N junction (Q-Photonics QEWLD-808). The epitaxial structure was grown on an n-type GaAs wafer and consisted of a 1500 nm n-type $Al_{0.55}Ga_{0.45}As$ layer, an undoped 200 nm $Al_{0.3}Ga_{0.7}As$ layer with a 10 nm GaAs quantum well in the center, a 1500 nm p-type $Al_{0.55}Ga_{0.45}As$ layer, and a 300 nm p-type GaAs contact layer. With current injection, the AlGaAs quantum well provides optical gain near $\lambda = 800$ nm. The layered structure results in vertical confinement of light in the undoped $Al_{0.3}Ga_{0.7}As$ layer via index guiding. The fundamental guided mode has the peak intensity at the location of the GaAs quantum well and experiences the highest gain.

### Laser Fabrication

The D cavities were fabricated by photolithography and wet etching. Once the cavity shape was defined by photolithography, electrical contacts consisting of 20 nm of Ti and 300 nm of Au were deposited by electron-beam evaporation. After lift-off, the metal contacts were used as etch masks and the wafer was wet etched to a depth of ~2 μm in a solution of phosphoric acid and hydrogen peroxide in water. This etching depth exceeds the waveguiding layer of $Al_{0.3}Ga_{0.7}As$, ensuring that light in the guided mode experienced a high refractive index contrast at the boundary of the D cavity due to the semiconductor-air interface. This high index contrast ensured strong optical confinement and formation of high-$Q$ resonances which experienced total internal reflection for a wide range of incident angles. Finally, a backside electrical contact consisting of 25 nm of Ni and 350 nm of AuGe was deposited via electron beam evaporation. For comparison, standard Fabry-Perot (FP) cavities in ridge waveguide geometry were fabricated on the same wafer.

### Optical Characterization

The experimental setup for lasing characterization was drawn schematically in Fig. 2(a). Electrical current was injected through the top contact with a Tungsten probe needle. The emission was then collected with a multimode fiber and directed to the entrance slit of a spectrometer (Acton Research Corporation Spectra Pro 300i). The emission spectrum was recorded as a function of pump current. The sample was mounted on a heat sink and operated at room temperature. To reduce thermal effects, pump current was turned on for 1 μs and repeated at a rate of 1 kHz. Since 1 μs was much longer than the response times of the material and the cavity, the system reached quasi-steady state.

**Acknowledgements**

B.R. and M.A.C. acknowledge support from the National Institutes of Health under Grant No. 1R21EB016163-01A1. A.D.S. and A.C. acknowledge support from National Science Foundation under Grant No. DMR-1307632. H.C. acknowledges support from the National Science Foundation under Grant No. ECCS-1128542 and the Office of Naval Research under Grant No. ONR MURI SP0001135605.